\newcommand\actaa{{Acta Astron.\,}}
\newcommand\apjl{{ApJ\,}}
\newcommand\apjs{{ApJS\,}}
\newcommand\aap{{A\&A\,}}
\newcommand\nat{{Nature\,}}

\newcommand{\Kepler}{\textit{Kepler}}

\documentclass[graybox, natbib, footinfo]{svmult}

\usepackage{natbib}       

\usepackage{mathptmx}       
\usepackage{helvet}         
\usepackage{courier}        
\usepackage{type1cm}        
\usepackage{makeidx}         
\usepackage{graphicx}        
\usepackage{multicol}        
\usepackage[bottom]{footmisc}
\usepackage{url}

\makeindex             

\begin{document}

\title*{Aspects of observational red giant population seismology}
\author{Joris De Ridder}
\institute{J. De Ridder \at Instituut voor Sterrenkunde, K.U.Leuven, Celestijnenlaan 200D, 3001 Leuven, Belgium, \email{joris@ster.kuleuven.ac.be}}
%
%
\maketitle

\abstract{The space missions CoRoT and \Kepler\ provide us with large samples of red giant stars wherein non-radial solar-like oscillations can be detected. This leads to the exciting opportunity to do population seismology. In this paper we give a short overview of some relevant technical aspects of the two satellites, we list and comment on some important target selection biases relevant for population seismology, and we make a case to use kernel density estimates as an alternative for histograms to characterize population characteristics.}

\section{Introduction}
\label{sec:1}

\emph{Population seismology} uses oscillation properties of a population of variables of a particular type to extract information about their stellar interior, or about the spatial structure, history, and future evolution of the population. In case of a homogenous population such as often assumed inside stellar clusters, population seismology is often referred to as \emph{ensemble seismology}. In case of a heterogenous population, such as the collection of red giants in a certain field of view, it is assumed that the population is sufficiently representative to extract information about our Galaxy as a whole.

Basically four ingredients are needed for population seismology: 1) seismic observables for a fairly large population, 2) a proper understanding of the selection biases that were involved in selecting the population members, 3) theoretical population models, and 4) a quantitative method to compare observations with theory to derive information. None of these ingredients are trivial and usually require an extensive amount of work. In this paper we will touch upon topics 1), 2), and 4), and we refer for the theoretical population models to the contribution of Miglio in this volume.

Requiring a set of seismic observables of a fairly-sized sample of variables implies gathering high-quality time series for a large number of stars. From the ground this turns out to be very challenging. The better large time-resolved databases are currently surveys such as OGLE \citep{Udalski97}, and ASAS \citep{Pojmanski02}. Unfortunately, these datasets are not good enough to detect the really low-amplitude solar-like oscillations in red giants. In addition, these surveys are done with small telescopes observing relatively bright and therefore relatively near giants. This limits the possibility to deeply probe our galaxy. 

Dedicated extensive spectroscopic campaigns of very bright giants (e.g.~\citealt{Frandsen02, DeRidder06}) show that it is possible to detect unambiguously low-amplitude solar-like oscillations in red giants. However, it was not until the arrival of the satellite CoRoT that it was possible to convincingly prove the existence of non-radial oscillations in giants \citep{DeRidder09}. Although CoRoT was capable of making a giant leap forward in red giant seismology, even this satellite is not sensitive enough to detect oscillations in all types of red giants. Giants with a frequency $\nu_{\rm max}$ (at maximum oscillation power) larger than 120 $\mu$Hz are often out of reach for CoRoT, which includes the rare but important low-mass low-luminosity giants. It took an even larger space telescope, \Kepler, to also detect this type of giants \citep{Bedding10}. Currently CoRoT and \Kepler\ are the only two instruments which allow red giant population seismology.

\section{CoRoT and \Kepler\ in a nutshell}
\label{sec:2}

In this section we give a brief overview of the technical aspects of CoRoT and \Kepler\ relevant for population seismology. 

The CoRoT spacecraft (see e.g.~\citealt{Baglin09}) was successfully launched on 27 December 2006 into a circular polar orbit around Earth. Using its 27 cm sized pupil, it was designed to continuously monitor the brightness of stars in a field of view of about $2.7^{\circ}\times3.05^{\circ}$ somewhere inside two circular parts of the sky, called \textit{the eyes of CoRoT} near the galactic plane at $\alpha\approx 6h50$ and $\alpha\approx18h50$. As the satellite orbits Earth, it is forced to regularly interrupt its observation runs to repoint in another direction to avoid the sunlight. This leads to two 150 run periods per year (so called \textit{long runs}, LRs), and two short runs of about 30 days (SRs). The first two long runs were chosen in the constellations of Monoceros and Aquila. From the beginning it was made clear that CoRoT had two goals: 1) finding exoplanets, and 2) asteroseismology. This led to a technical design of 4 CCDs of 2046$\times$4096 pixels each: 2 seismofield CCDs (A1 and A2) and 2 exoplanet field CCDS (E1 and E2). The former are meant to observe relatively few but bright targets of asteroseismological interest, while the latter aim to observe together about 12000 relatively faint targets to detect exoplanet transits. The exoplanet field targets have  visual magnitudes between 12 and 16, and are observed with an exposure time of 512s. A significant fraction of these targets are red giants, which make up a population excellently suited for population seismology. More information about the CoRoT mission can be found in ESA SP-1306 (2006). 

NASA's \Kepler\ mission (see e.g.~\citealt{Borucki09}) was launched almost 3 years after CoRoT on 7 March 2009. The spacecraft is in a heliocentric orbit which allows it to stare for at least 3.5 years at the same field of view in the Cygnus constellation. It's aperture is about 95 cm, and it features 42 CCDs covering about 105 square degrees of the sky. Roughly 130,000 targets within a dynamic magnitude range of $9 \le m_{V} \le 16$, are continuously monitored, and also here a significant fraction turn out to be red giants, which are observed with an exposure time of about 30 min (long cadence) or of about 1 min (short cadence).

How do these satellites compare for a population seismologist? CoRoT had the advantage of being first, which led to the first population seismology paper relying on non-radial oscillations in red giants observed by CoRoT \citep{Miglio09}. However, the performance of \Kepler\ in terms of S/N or in terms of the length of the time series goes beyond anything what CoRoT can provide. This allows for more accurate seismic observables, and an evolutionary stage coverage that includes red giants with $\nu_{\rm max} \ge 120\ \mu{\rm Hz}$ (cf.~Sect.~\ref{sec:1}). CoRoT, on the other hand, changes its field of view regularly, which allows to probe different directions in our galaxy to study, for example, possible spatial metallicity gradients. In this sense, the two missions are complementary.

\section{Selection biases}
\label{sec:3}

Comparing observations with theory usually involves comparing observational and theoretical histograms (or equivalent) of some seismic observable. A difficult aspect of population seismology is to investigate if there are any selection biases that may have caused some of the bins to be less populated than they should be. Failing to identify these selection biases, or failing to make a proper impact analysis of them will likely result into an erroneous interpretation of the data.

There can be basically four reasons why a red giant was selected to be observed by CoRoT or \Kepler. First, by selection on scientific grounds by, for example, the red giant analysis team. Secondly, simply because there were roughly the same number of targets in the field as available slots on the CCDs, so that every target is observed. Thirdly, failure to avoid red giants by planet hunters. And fourthly, as astrometric targets. Each of these reasons may have different (although overlapping) selection criteria and biases.

In what follows we list and comment on some of the common selection criteria.
\begin{itemize}
\item The most obvious selection criteria are based on observationally derived stellar parameters. For example, every \Kepler\ astrometric giant satisfies the criteria $T_{\rm eff} < 5400\ {\rm K}$, $R/R_{\odot} > 2$, and $\log g < 3.8$. It's important to realize, however, that the derivation of stellar parameters for the databases of targets of CoRoT and \Kepler\ are for the largest part not based on spectroscopic data, but rather on fitting a small set of photometric measurements. As a result, both the parameters and their quoted uncertainties are not always reliable, especially for the hotter and the cooler stars. To what extend this affects a histogram of the observed giant population remains to be investigated, however.   
\item The allowed magnitude range of the targets is determined by the dynamic range of the instrument, but may be limited for specific purposes. The astrometric giants of the \Kepler\ mission, for example, have a rather narrow magnitude range of 11 to 12.5. The fainter the target, the lower the S/N ratio will be, and the more difficult it will be to extract useful asteroseismic information. While this is unlikely an issue for \Kepler, it could be for CoRoT, depending on what type of asteroseismic information one is looking for. As the signal strength may vary because of the stochastic nature of the oscillations, the upper magnitude boundary of the sample of useful red giants can be rather fuzzy, in contrast with the lower magnitude boundary which is set to avoid saturation.  
\item It is not only the saturation risk of a potential red giant target that matters. Even if the target is sufficiently faint, it may still be rejected if a neighboring star on the CCD would likely saturate and bleed upon it.
\item Red giants suffering too much from crowding (i.e. from background objects contributing more than e.g. 5\% of the flux in the mask) are rejected to avoid a decrease in S/N, a signal contamination, or inaccurate centroiding in the case of astrometric giants.
\item Astrometric giants also require a small parallax and a small proper motion, for obvious reasons.
\item The effective field of view (FOV) is a bit smaller than the quoted instrumental FOV. The reason is that the silicon edge is usually avoided for target selection. This is especially the case for \Kepler, where the exact position on the CCD depends on the season, and where one requires that a giant star is observable in all seasons.
\item The appearance of instrumental peaks in the power spectrum can have a large effect on the final red giant sample. In the case of the CoRoT sample, for example, there appear to be hardly any giants with a a $\nu_{\rm max} \ge 120\ \mu{\rm Hz}$ for which reliable seismic observables can be derived. This is not only due to S/N limitations, but also due to the occurrence of peaks coming from a mixture of orbital and diurnal frequencies at $(161.7 \pm k\cdot 11.6)\  \mu{\rm Hz}$, with $k$ an integer \citep{Mosser10} which make the data reduction and analysis a lot more difficult.
\item The limited number of available slots on the CCD can also introduce a selection bias for specific types of giants. In the case of \Kepler\ the long cadence (LC) time series have a Nyquist frequency of about $\nu_{\rm Nyq} \approx 280\ \mu{\rm Hz}$. Red giants with a $\nu_{\rm max} \ge \nu_{\rm Nyq}$ therefore need to be observed in short cadence (SC). The SC slots are, however, mainly used for main sequence stars and subgiants to fulfill the space mission's main science objective. The number of low-luminosity low-mass giants will therefore be underrepresented in the \Kepler\ red giant sample. Although much less, there is also an over-demand for the LC slots. Usually a selection is made to ensure a large variety of giants ($\nu_{\rm max}$, magnitude, metallicity, ...).
\end{itemize}

The list above is not exhaustive, and it should be clear that it would be rather challenging (if not impossible) for a theoretician to reproduce all selection criteria to produce a suitable synthetic population. Some of the criteria may not even be formulated in the form of hard constraints. However, one usually does know the observational data that were available to make the selection. A better alternative may therefore consist of Monte Carlo resampling the corresponding synthetic data to produce an optimal distribution of synthetic populations with the observed characteristics. All statistical inferences can then be made in a Bayesian way which makes it easy to include this distribution.

\section{Comparing population characteristics}

Visualizing seismic population characteristics is often done using a histogram (e.g. \citealt{Miglio09}). In this section, we would like to advocate using a better alternative: kernel density estimates (KDE). Histograms depend on two free parameters both of which can have a rather large impact on its shape and appearance: the origin, and the bin width $h$. Varying these parameter may change the appearance of a peak in the histogram from symmetric to skewed, or vice versa, and may cause small features to appear or disappear. The problem of the histogram origin can in principle be solved by averaging histograms made with different bin grids. Choosing a good bin width, however, remains difficult if the shape of the true distribution is unknown. If we choose the bin width too small, we're under-smoothing and suffer a large variance. A too large bin width, on the other hand, leads to over-smoothing and a large bias.

A useful alternative would be a non-parametric density estimate using a kernel function $K$. Given a set of $n$ observed values $\{X_{i}\}$, the probability density function $f(x)$ of the corresponding quantity is estimated by
\begin{equation}
\hat{f}_{h}(x) = \frac{1}{n h} \sum_{i=1}^{n} K\left(\frac{x-X_{i}}{h}\right)
\end{equation}
where $K$ is a kernel function and $h$ the bandwidth. Often used Kernels are the Gaussian kernel $K(u) \equiv exp(-u^{2}/2)/\sqrt{2\pi}$ or the Epanechnikov kernel $K(u) \equiv 0.75\ (1-u^{2})\ I(|u| \le 1)$ where $I$ is the indicator function. 

\begin{table}
\caption{Asymptotic properties of the histogram and the kernel density estimator (KDE). Using the mean integrated squared error (MISE) as a measure, the KDE proves to have a better convergence rate compared to the histogram.}
\label{tab:1}       
%
%
\begin{center}
\begin{tabular}{p{2.2cm}p{2.1cm}p{2.1cm}p{2.1cm}p{2.1cm}}
\hline\noalign{\smallskip}
       & Bias & Variance & $h_{\rm opt}$ (MISE) & MISE $(f_{h})$ \\
       & $(h\rightarrow 0)$ & $(nh\rightarrow\infty)$ & $(nh\rightarrow\infty)$ &  $(nh\rightarrow\infty)$ \\ 
\noalign{\smallskip}\svhline\noalign{\smallskip}
Histogram & $\sim h$  & $\sim (nh)^{-1}$ & $\sim n^{-1/3}$ & $\sim n^{-2/3}$ \\
Kernel & $\sim h^{2}$ & $\sim (nh)^{-1}$ & $\sim n^{-1/5}$ & $\sim n^{-4/5}$ \\
\noalign{\smallskip}\hline\noalign{\smallskip}
\end{tabular}
\end{center}
\end{table}

\begin{figure}[t]
\sidecaption
\includegraphics[scale=.5]{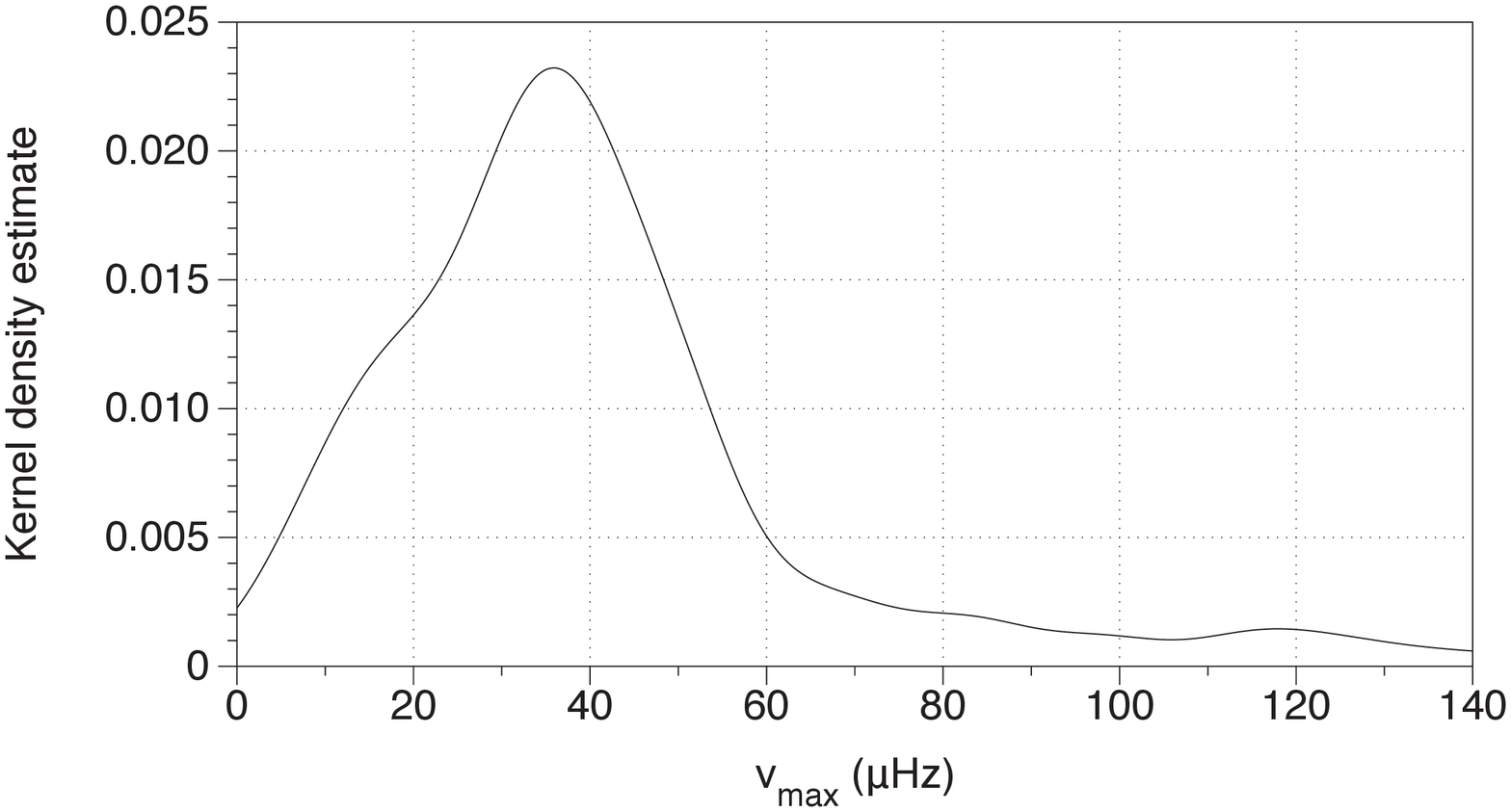}
\caption{Kernel density estimate of the distribution of the frequency $\nu_{\rm max}$ of maximum power of a sample of CoRoT giants in the first anti-center field (LRa1). Seismic data from \citet{Mosser10}.}
\label{fig:1} 
\end{figure}
\begin{figure}[h]
\sidecaption
\includegraphics[scale=.5]{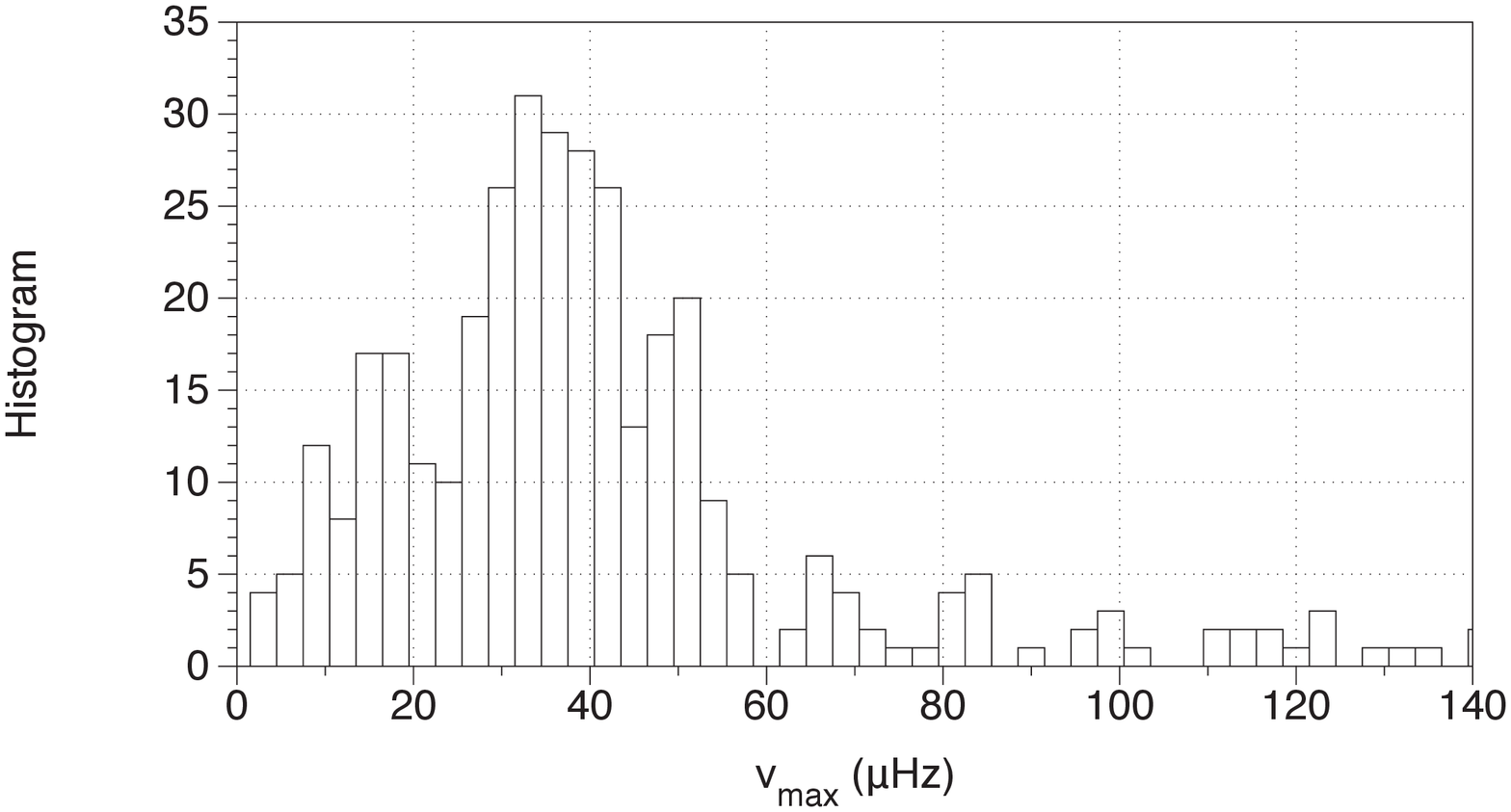}
\caption{Kernel density estimate of the distribution of the frequency $\nu_{\rm max}$ of maximum power of a sample of CoRoT giants in the first anti-center field (LRa1). Seismic data from \citet{Mosser10}.}
\label{fig:2} 
\end{figure}

At first sight, not much seems to be gained. There is still a bandwidth $h$ to choose, which can be seen as the analogue of the bin width of a histogram, and now we also have to choose which kernel is best suited. In practice, however, the latter is not a problem because it usually makes very little difference which kernel is exactly used. An important reason to choose KDEs is that they have better asymptotic properties than a histogram. Quantifying the performance of a density esimator can be done by looking at the bias ${\rm E}[\hat{f}_{h}(x)] - f(x)$, the variance ${\rm Var}[\hat{f}_{h}(x)]$, and the mean integrated squared error (MISE):
\begin{equation}
{\rm MISE}[\hat{f}_{h}] \equiv E\left[\int\limits_{-\infty}^{+\infty} (\hat{f}_{h}(x) - f(x))^{2}\ dx \right]
\end{equation}
The latter quantity can used to derive the optimal bandwidth $h_{\rm opt}$ that minimizes the MISE. Table \ref{tab:1} (adapted from \citealt{Hardle90}) compares the results for histograms and KDEs. Clearly, both the bias and the MISE of the KDE have a higher convergence rate than those of histograms.

Although the optimal bandwidth $h_{\rm opt}$ derived from the MISE is handy to derive theoretical results, it has no practical use if the shape of the true distribution is not known. It can be shown, however (see e.g.~\citealt{Hardle04}) that using the integrated squared error (ISE)
\begin{equation}
{\rm ISE}[\hat{f}_{h}] \equiv \int\limits_{-\infty}^{+\infty} (\hat{f}_{h}(x) - f(x))^{2}\ dx
\end{equation}
together with cross-validation can be used to compute an optimal value for the bandwidth. This method, as well as alternative statistically justified methods are readily available in the statistical package R. In addition, we refer to \citet{Hardle04} on how to compute confidence intervals and confidence bands for the KDEs, something which is not easily available for histograms.
As a final illustration, we show the KDE of the frequency of maximum power of a sample of CoRoT giants in the LRa1 field in Fig.~\ref{fig:1}, and a histogram of the same sample in Fig.~\ref{fig:2}. We hope this short introduction will convince population seismologists of the benefits of using KDEs.



\end{document}